# Chromatin Folding in Relation to Human Genome Function

Julio Mateos-Langerak[1,5], Osdilly Giromus[1], Wim de Leeuw[2], Manfred Bohn[3], Pernette J. Verschure[1], Gregor Kreth[4], Dieter W. Heermann[3], Roel van Driel[1] and Sandra Goetze[1,6]

[1] Swammerdam Institute for Life Sciences, University of Amsterdam, BioCentrum Amsterdam, Kruislaan 318, 1098 SM Amsterdam, The Netherlands

[2] National Research Institute for Mathematics and Computer Science, Kruislaan 413, 1098 SJ Amsterdam, The Netherlands

[3] Institute of Theoretical Physics, University of Heidelberg, Philosophenweg 19, 69120 Heidelberg, Germany

[4] Kirchhoff Institute for Physics, University of Heidelberg, INF 227, 69120 Heidelberg, Germany

[5] present address: Institute of Human Genetics, Centre National de la Recherche Scientifique, rue de la Cardonille 141, 34396 Montpellier, France

[6] present address: Institute of Molecular Biology, University of Zurich, Winterthurerstrasse 190, 8057 Zurich, Switzerland

**Summary**

Three-dimensional (3D) chromatin structure is closely related to genome function, in particular transcription. However, the folding path of the chromatin fiber in the interphase nucleus is unknown. Here, we systematically measured the 3D physical distance between pairwise labeled genomic positions in gene-dense, highly transcribed domains and gene-poor less active areas on chromosomes 1 and 11 in G1 nuclei of human primary fibroblasts, using fluorescence in situ hybridization. Interpretation of our results and those published by others, based on polymer physics, shows that the folding of the chromatin fiber can be described as a polymer in a globular state (GS), maintained by intra-polymer attractive interactions that counteract self-avoidance forces. The GS polymer model is able to describe chromatin folding in as well the highly expressed domains as the lowly expressed ones, indicating that they differ in Kuhn length and chromatin compaction. Each type of genomic domain constitutes an ensemble of relatively compact globular folding states, resulting in a considerable cell-to-cell variation between otherwise identical cells. We present evidence for different polymer folding regimes of the chromatin fiber on the length scale of a few mega base pairs and on that of complete chromosome arms (several tens of Mb). Our results present a novel view on the folding of the chromatin fiber in interphase and open the possibility to explore the nature of the intra-chromatin fiber interactions.

**Introduction**

The chromatin fiber inside the interphase nucleus of higher eukaryotic cells is folded in a hierarchical manner. The basic filament is formed by wrapping DNA around a histone protein octamer, forming a nucleosomal unit every 180 to 250 bp, the precise number depending on the length of the linker DNA (1, 2). The next folding level is an about 30 nm diameter fiber in which the nucleosomes are probably arranged in a zigzag manner (2, 3). Remarkably little is known about higher order levels of chromatin folding (4). Various studies have presented evidence that chromatin is organized in loops, which may be attached via scaffold/matrix-attachment regions (S/MARs) at their bases to a still elusive structure that is called nuclear scaffold/matrix (5-7). Recent investigations indicate that chromatin loops are dynamic, relate to gene activity (8), and involve specific proteins, including the nuclear matrix protein SatB1 (9, 10) and the insulator binding protein CTCF (11, 12). Other studies link chromatin loops to the formation of transcription factories in which genes from different positions on a chromosome and possibly from different chromosomes come together (11, 13-16). In addition, there is considerable evidence by light microscopy that large genomic loci (typically several Mb) constitute subchromosomal loop domains that may extend away from chromosome territories after activation (17-19). Transcriptional activation has been shown to result in chromatin decondensation, which can be visualized by light microscopy. Examples in mammalian cells of decondensation and looping are the MHC locus (17) and the HoxB and D gene cluster (19-21). Many more gene clusters probably exist that behave in similar ways (18, 22). Together, these studies indicate that there is a tight correlation between folding of the chromatin fiber and its transcriptional activity.

Despite this relationship, our understanding of how the chromatin fiber is spatially organized is remarkably limited. Imaging techniques do not allow one to follow the folding path of the filament in the interphase nucleus (4, 23). Therefore, indirect approaches are being used to obtain information about 3D chromatin structure. Several groups have employed fluorescence in situ hybridization (FISH) to analyze chromatin folding by establishing the relationship between the physical distance R (in μm) between genomic sequence elements in interphase and the genomic distance g (in Mb). Polymer physics can be used to extract information about the folding of the chromatin fiber inside the interphase nucleus (24-26). Doing so, experimental results have been interpreted in various ways, including random walk (RW) polymer model (27, 28), large (2 to 3 Mb) loops attached to a randomly folded backbone (random walk giant loop folding) (25, 27), and the related organization in about one Mb-size rosette-like structures, containing 120 kb loops, linked by a short flexible part of the chromatin fiber (multi-loop subcompartment folding) (29).

Here, we use a similar approach by establishing the precise relationship between physical and genomic distances to obtain quantitative information about chromatin folding of non-repetitive domains in the human genome. In this study we analyzed two types of genomic areas: (i) gene-dense and transcriptionally highly active domains and (ii) domains that are gene-sparse and lowly expressed. Such domains are defined in the human transcriptome map (30, 31). Recently, Goetze et al. (32) have shown that such domains, named ridges (regions of increased gene expression) and anti-ridges respectively, differ in compaction of the chromatin fiber, in shape, and in position inside the nucleus. Here, we extend these studies by analyzing the 3D folding path of the chromatin fiber in ridges and anti-ridges, thereby directly relating chromatin folding with genome function. Using FISH under conditions that optimally preserve nuclear structure in combination with 3D imaging and image analysis, we compare 3 to 12 Mb genomic domains located on the q-arm of chromosomes 1 and 11, which are either rich in highly active genes, or genes-sparse and transcribed at low levels. Also, we investigated more extended genomic stretches (30 to 75 Mb) spanning most of the q-arms of chromosomes 1 and 11. Interpretation of our data in terms of polymer physics imposes stringent limits to models that describe the folding of the chromatin fiber. We show that the RW folding model for chromatin cannot describe the results of the in situ measurements satisfactorily, in contrast to what has been concluded by others. Rather, the chromatin fiber appears to exist in an ensemble of related relatively compact globular states (GS) that are maintained by the interplay between self-avoiding forces and intra-chain interactions, related to loop formation. Chromatin of gene-rich and of gene-poor domains differs in the degree of compaction. Moreover, we present evidence for two different chromatin folding regimes, one at the length scale of subchromosomal domains, i.e. a few Mb, the other at the length scale of many tens of Mb, i.e. the level of complete chromosome arms.

**Results**

*Experimental approach*

The aim of this study was to analyze and compare the folding of the chromatin fiber in two functionally different types of chromatin domains in the human genome, i.e. gene-rich domains that show a high gene expression (ridges) and gene-poor areas that are transcribed at a low level (anti-ridges) (30, 31). The human transcriptome map shows that our genome contains several tens of such gene-dense and gene-poor areas of 3 to 15 Mb on various chromosomes (30, 31). Using FISH under structure-preserving conditions in combination with automated 3D image acquisition and 3D image analysis, we set out to establish the quantitative relationship between the physical distance R and the genomic distance g within ridges and anti-ridges on the q-arms of chromosomes 1 and 11. Ridges and anti-ridges on these chromosomes are particularly pronounced. To analyze chromatin folding, 60 bacterial artificial chromosomes (BACs, about 165 kb per BAC) were selected that recognize approximately equally spaced sequences, together spanning a large part of the q-arm of chromosome 1 and about the complete q-arm of 11 (see supporting information (SI) Table 1). For each 3D distance measurement at least 50 nuclei were imaged and quantitatively evaluated, resulting in practice in over 80 measurements per distance measurement between a pair of BAC probes. The relationship between R and g was analyzed in human primary fibroblasts exclusively in the G1 cell cycle phase to reduce cell cycle effects on chromatin folding. Fig. 1A shows the 1q and 11q areas of the human transcriptome map that have been analyzed. The beginning of the arrow above the map indicates the position of the reference FISH probe, whereas the arrow head marks the location of the FISH probe that has the largest genomic distance to the reference probe. All distances have been determined with respect to the reference probe. Green arrows and green data points refer to ridges, red ones to anti-ridges. Black arrows indicate long distance measurements beyond ridge and anti-ridge domains. Distances were measured in 3D space between the centers of gravity of the 3D FISH signals of the individual BAC probes. All experiments were carried out on normal primary human fibroblasts in the G1 phase of the cell cycle.

*Distance measurements in ridge and anti-ridge domains*

Fig. 1B shows distance plots (R vs. g) of the 3 Mb ridge and anti-ridge domains on chromosome 1q and the 10 -12 Mb ridge and anti-ridge domains on chromosome 11q.

Above a few Mb genomic distance the increase in average 3D physical distance levels off. Average R values for anti-ridges are lower than found for ridges, reflecting the different degrees of compaction of these domains (32). A strikingly large cell-to-cell variation is observed for all distance measurements. This is not due to errors in 3D measurements, which are better than 50 nm (see Materials and Methods). Also differences between cells due to different cell cycle stages are unlikely, because all analyzed nuclei were in G1. Apparently, cell-to-cell variation is an intrinsic property of chromatin folding, indicating that, rather than folding in a unique 3D configuration chromatin folding constitutes a conformational ensemble.

*Distance measurements beyond ridge and anti-ridge domains*
Distance plots covering a large part of the q-arm of chromosome 1 (27 Mb) and essentially the complete q-arm of chromosome 11 (75 Mb) are shown in Fig. 1C. Remarkably, the physical distance R levels off at genomic distances beyond 5 to 10 Mb and does not increases significantly at distances up to 75 Mb. These results suggest that the chromatin fiber of the 1q and 11q chromosome arms is confined to a limited volume. The leveling off of distances is in the μm range, similar to the size of chromosome territories, suggesting that this phenomenon is related to the confined space that chromosomes occupy in interphase nuclei (33).

*Polymer models of chromatin folding of ridges and anti-ridges*
If it is assumed that the chromatin fiber can be modeled as a polymer, then such model would predict the relationship between the average mean square spatial distance $<R^2>$ between two markers along the fiber and the polymer contour length, i.e. the effective genomic length $N_m$ of the chromatin fiber. $N_m$ is related to the genomic distance g by the Kuhn segment length b ($g = bN_m$). The Kuhn length is connected to the persistence length P by $P = b/2$. The average mean square spatial distance $<R^2>$ is used because the polymer forms an ensemble of conformations in space over which the model averages. Whatever polymer model is chosen, one can expand the dependence of the spatial distance R on the contour length in terms of a non-trivial power series in the contour length.

$$<R^2> = b^2 N_m^{2\nu} ( 1 + a_1 N_m^{b_1 - 2\nu} + a_2 N_m^{b_2 - 2\nu} + ... ) \qquad (1)$$

Here b is the Kuhn length, the parameters $a_i$ are the coefficients of the expansion and $b_i$ the fractional exponents in the expansion. The value of the exponent ν differs for different polymer models. Three basic polymer models are relevant for modeling the chromatin fiber. Each polymer model is characterized by a specific value of the exponent ν.

(i)     random walk (RW) model: ν = 0.5

(ii)    self-avoiding walk (SAW) model: ν = 0.588

(ii)    globular state (GS) model: ν = 1/3

The RW model ignores the excluded volume of the polymer and assumes that there are no intra-chain attractive forces. In this case the values $a_i$ are all zero. The SAW model takes the excluded volume into account and the expansion asymptotically converges to $b^2 N_m^{2\nu}$ for large values of $N_m$. The GS model assumes intra-polymer attractive interactions that counteract self-avoidance forces, resulting in a collapsed globular state of the polymer if the attractive interactions dominate (34). Here too the expansion asymptotically converges to $b^2 N_m^{2\nu}$ at large $N_m$. While for the three models (RW, SAW and GS) the $a_i$ values are all zero for large $N_m$, the above expansion (Eq. 1) takes into account deviations of the leading order behavior, due to for example loop formation.

To conduct a highly sensitive comparison between these three polymer models we divided out the leading order term $N_m^{2\nu}$ of Eq. 1 and analyzed the ratio $<R^2>/N_m^{2\nu}$ as a function of the contour length $N_m$ for the experimental data set shown in Fig. 1B. Figures 2A-C show the result for the three different ν values: ν=0.5 (RW), 0.588 (SAW) and 1/3 (GS) for the data sets of the ridge and anti-ridge of chromosome arm 1q. Each of the three models predicts that the ratio $<R^2>/N_m^{2\nu}$ is independent of the genomic distance. Fig. 2 shows that this condition is only fulfilled if the experimental data is interpreted in terms of the GS model, i.e. for ν = 1/3. For other values of ν, typical for the RW and SAW models, this is not the case. Evidently, the folding of the chromatin fiber can best be described with the GS model, making the RW and SAW polymer models less likely. The GS polymer model fits the ridge and the anti-ridge data sets equally well, indicating a GS folding for both, the only difference being the polymer parameters.

Yokota et al. (27) have carried out similar FISH-based distance measurements on a telomeric segment of chromosome 4 of cultured human fibroblasts, which can be identified as a ridge according to the HTM. We have incorporated their data set in our analyses in Figs. 2A-C. Note that Yokota et al. (27) carried out distance measurements in 2D, rather than 3D. For the analyses in Fig. 2 this results in a difference in scaling compared to our 3D data set. Fig. 2A-C shows that also the results of Yokota et al. (27) can be described best by a GS polymer model, rather than the RW model that was by themselves.

Analyzing the long distance measurements of Fig.1C in terms of the GS polymer model, suggests that different folding regimes exist for different genomic length scales. In Fig. 2D this is seen by the initial high value for $<R^2>/N_m^{2\nu}$, reflecting the more open globular state of the local ridge-type of chromatin folding near the reference point in the distance measurements (up to a few Mb). At longer distances, i.e. above about 20 Mb, chromatin seems to fold in a more compact globular state, shown by the lower value of $<R^2>/N_m^{2\nu}$. This is conceptually similar to the two-level random walk/giant loop folding model of Sachs et al. (25), although they interpreted their results in terms of a RW polymer model rather than a GS model.

**Discussion**

We have compared the folding of the chromatin fiber of two gene-rich and highly expressed genomic areas with that of two lowly expressed and gene-poor domains on the q-arms of chromosomes 1 and 11 in primary human fibroblasts. These genomic domains, named ridges and anti-ridges respectively, were selected from the human transcriptome map (30, 31). Using semi-high throughput FISH on structurally preserved cells, in combination with 3D image analysis, we established the relationship between the physical distance R and the genomic distance g of the chromatin fiber. Results in Fig. 1 show that chromatin folding in ridges is significantly different from that in anti-ridges, reflecting the different functional states of the domains. Our measurements show that the 3D distance between pairs of FISH probes increases only up to a genomic distance of a few Mb. At this genomic scale our results qualitatively agree with earlier studies (24, 25, 27, 35). At larger genomic distances the physical distance does not depend much on genomic length. This leveling off occurs at larger distances for ridges than for anti-ridges (Fig. 1B), which is consistent with previous measurements showing that anti-ridges are about 30% more compact (smaller volume per Mb) and have a more spherical shape than ridges (32). Leveling off observed at large distances around 2 μm (Fig. 1C) is probably related to the confinement of interphase chromosomes in chromosome territories (33).

The basic polymer models that we explore in this study are described by two main parameters, i.e. the number of base pairs k that equals one chain segment and the chain physical segment length l, both describing the Kuhn length (which is two times the persistence length). Fitting these parameters to the GS model yields values for the ratio $l^2/k^{2\nu}$, which has the dimension $nm^2/bp^{2/3}$. Bystricky et al. (36) have estimated the persistence length of the chromatin fiber of budding yeast: 197 +/- 62nm (Kuhn length 394 +/- 124 nm). As a first approximation, one may assume that human chromatin of ridges and of anti-ridges has a similar Kuhn length as found for yeast. If so, one can calculate from the results in Fig. 2C that the corresponding genomic length is about 24 kb for ridges and 380 kb for anti-ridges, leading to the conclusion that the chromatin compaction rate in the order of 0.1 and 1 Mb/μm, respectively. This suggests that the average chromatin-packing ratio in anti-ridges is one order of magnitude higher than that of ridges. However, these values have to be treated with caution, because we do not have independent measurements of the Kuhn length of chromatin in ridges and anti-ridges.

Yokota et al. (37) have compared chromatin structure of R and G band chromatin of several chromosomes in human fibroblasts, using a similar approach. They interpreted their results in terms of a random walk polymer model, the G-band being somewhat more compact than the R-band. Although there is some correlation between the Giemsa banding pattern and ridges and anti-ridges in the human transcriptome map (SI Table 2), this relationship is limited (Fig. 1 A). For instance, the regions of chromosome 1 studied here are pronounced examples of a ridge and an anti-ridge. Nevertheless, they are both part of a G-band. Obviously, the human transcriptome map is a better representation of local genome activity than the Giemsa banding pattern.

Recently, the Bickmore group has biochemically fractionated chromatin fragments on the basis of differences in compactness (38). Fractions enriched in open chromatin correlate well with ridges in the HTM. Using a similar approach as used by us, they showed that the ridge on distal 11p15.5 indeed is relatively decondensed, in agreement with our findings.

Our results differ from those of others that concluded that chromatin behaves as a random-walk polymer at distances up to several Mb (25, 27, 29). Results from longer distance measurements have been taken as evidence for the existence of rosette-like chromatin domains, consisting of multiple chromatin loops, which are interconnected by stretches of flexible chromatin fibre (27, 29). Our analysis in terms of polymer physics is highly sensitive in comparing the experimental data sets with predictions of the three basic polymer models: the random walk (RW) model, the self-avoiding walk (SAW) model, and the globular state (GS) model. We show that only the GS polymer correctly predicts that for ridges and anti-ridges the measured ratio $<R^2>/N_m^{2\nu}$ is independent of the contour length, i.e. the genomic distance (Figs. 2A-C). Our results also differ from those of others that observe a monotonous increase of R over genomic distances up to 180 Mb and interpret results as evidence for a RW polymer model (25, 27, 28, 35). Reasons for such a discrepancy might be differences in sample preparation. Most of the earlier experiments include repeated freeze-thawing and acid-methanol fixation steps, which may not sufficiently preserve 3D chromatin structure. Also, these studies measure distances in 2D projection rather than in 3D.

In the course of our study we observed a cell-to-cell variation of measured distances R in the order of a micrometer (Figs. 1B, C). This variation appears to be an intrinsic property of interphase chromatin folding. Cell cycle effects can largely be ruled out, since all measurements have been made in G1 fibroblasts. Also, we show that the accuracy of measuring 3D distances is better than 50 nm and therefore measuring errors can be excluded as a main cause of cell-to-cell variation. The measured variation may in part reflect local constrained diffusion of the chromatin fiber, which is in the submicron range and has been observed in budding yeast (39, 40) and in mammalian cells (41-43). Evidently, the chromatin fiber of a ridge or an anti-ridge does not follow one unique 3D path. Rather, an ensemble of different chromatin fiber conformations exists in otherwise identical cells, each conformation being compatible with proper cell function under the given experimental conditions. Our measurements show that the folding ensembles of ridges and of anti-ridges are significantly different, most likely reflecting their different functional states.

The GS polymer model for ridge and anti-ridge chromatin folding suggests that intra-chromatin fiber crosslinks exist. These may be related to loop formation, including for instance the formation of transcription factories (11, 13-16) and/or binding to a putative nuclear scaffold (5, 6, 25). This is true for ridge and anti-ridge chromatin, which has been analyzed in this study, as well as for the data of Yokota et al. (27). Analysis of the long distance data sets in Fig. 1C and by Yokota et al. (27) in terms of the GS polymer model (Fig. 2D) suggests a two-level chromatin folding regime. At short distances (up to a few Mb) the local folding state predominates, i.e. the local folding state of ridges or anti-ridges. At long distances, a folding regime prevails, irrespective of gene density or transcriptional activity. This folding state is typified by $<R^2>/N_m^{2v}$ values similar to those of what is observed for short distances in anti-ridges. For ridges that implies that at short genomic distances (few Mb) the value of $<R^2>/N_m^{2v}$ in Fig. 2D is the same as observed for ridges in Fig. 2C. Above about 20 Mb the value of the leading order drops to a lower value, comparable to what is found for anti-ridges (Fig 2C). Obviously different folding regimes exist for the chromatin fiber at different length scales.

Summarizing, our analysis of chromatin folding in human primary fibroblasts in G1 interphase shows that the folding path can best be described in terms of a globular state (GS) polymer model, i.e. a polymer that is spatially constrained by intra-fiber interactions that counteract self-avoiding repulsive forces. The GS model describes our data and those of the Trask group (27) considerably better than the RW model and the SAW model. This has important consequences for our thinking about chromatin folding in interphase. The GS model predicts that a considerable number of intra-chain interactions exist. We can only speculate about the molecular nature and the number of these interactions. They may be related to, for instance, enhancer looping (12), transcription factories (6) and S/MAR binding proteins, such as SatB1, which is implicated in chromatin loop formation (7, 9). The GS model applies equally well to the gene-dense and highly expressed ridges and the gene-sparse lowly expressed anti-ridge domains, differing quantitatively in the chromatin packing ratio. It would be of interest to extract better information about typical length scales over which the chromatin fiber can be regarded as stiff, i.e. the persistence length or the Kuhn length. For this, measurements are needed on length scales where the fiber can be considered as a stiff rod, as has been done in yeast (36). Also, to further specify polymer models of human interphase chromatin, a more detailed statistical analysis is of importance. Especially the distance distributions and higher-order moments of the measured data may give important hints. For such statistical analysis, however, 10 to 100 times more data points per distance measurements are required. The quantitative analysis of chromatin folding in the context of polymer models as presented here in combination with the manipulation of chromatin proteins, epigenetic state and gene activity, i.e. components that are likely to affect model parameters, will allow a systematic analysis of the contribution of each of these features to chromatin structure. Exploiting polymer models therefore provides a promising route towards understanding the molecular basis of chromatin folding.

**Materials and Methods**

*Cell culture and fluorescence in situ hybridization*

Human primary female fibroblasts (04-147) were cultured in DMEM containing 10% fetal calf serum, 20 mM glutamine, 60 μg/ml penicillin and 100 μg/ml streptomycin. Primary fibroblasts were used up to passage 25 to avoid effects related to senescence.

BACs were selected from the BAC-clones available in the RP11-collection at the Sanger Institute (SI Table 1). All BACs were end-sequenced to confirm their identity. Genomic distances were defined as the distance between de centers of the BACs. BAC DNA was isolated using the Qiagen REAL prep 96 kit (Qiagen, Venlo, The Netherlands) and DOP-PCR amplified. Nick-translation was used to label the probes, either with digoxigenin or biotin (Roche Molecular Biochemicals, Basel, Switzerland). FISH was carried out as described elsewhere with slight modifications (32). In short, cells in interphase were incubated with a 30 min pulse of 25 μM BrdU (Sigma-Aldrich, MO, USA) to label replicating DNA prior to fixation in 4% (w/v) paraformaldehyde, in order to detect S-phase cells. Denaturation was carried out at 78°C for 2 min in 2x SSC, containing 50% formamide. Hybridization was allowed to proceed overnight at 37°C. Post-hybridization washes were performed with 2x SSC/ 50% formamide at 45°C. All incubations for probe detection were carried out at room temperature in 4x SSC, containing 5% (w/v) non-fat dried milk. FITC-conjugated antibodies (Jackson Immunoresearch Laboratories, Inc., West Grove, PA, USA) and Cy3-conjugated streptavidin (Vecta Laboratories, Burlingame, CA, USA) were used to visualize the signals. BrdU was visualized using a primary anti-BrdU antibody from Roche (Roche Molecular Biochemicals, Basel, Switzerland). Slides were mounted in Vectashield (Vecta) containing 1 μg/ml DAPI (Vector Laboratories, Burlingame, CA, USA).

*Confocal laser-scanning microscopy*

For each experiment over 50 nuclei were imaged. Twelve-bit 3D images were recorded in multi-track mode to avoid crosstalk. We used an LSM 510 confocal laser-scanning microscope (Carl Zeiss, Göttingen, Germany) equipped with a 63x/1.4 NA Apochromat objective, using an Ar-ion laser at 364 nm, an Ar laser at 488 nm and a He/Ne laser at 543 nm to excite DAPI, FITC and Cy3, respectively. Fluorescence was detected with the following bandpass filters: 385–470-nm (DAPI), 505–530-nm (FITC) and 560-615 (Cy3). Images were scanned with a voxel size of 50x50x100 nm.

*Image processing and data evaluation*

Automated image analysis was carried out on raw data sets with the ARGOS software (44) to identify nuclear sites labeled by BACs and to compute their 3D position in the nucleus. Chromatic aberration was measured via Tetraspeck Microspheres (Molecular Probes, Eugene, OR, USA) and corrected for in the analysis. After background subtraction, images were treated with a bandpass filter to remove noise. Subsequently, images were segmented and ensembles of interconnected voxels were regarded as the site labeled by a BAC. The centre of mass was calculated for each labeled site at subvoxel resolution and 3D distances between BACS was measured. To estimate our systematic measuring error we hybridized cells with a mixture of the same BAC marked with two different fluorophores and measured the distances between the two signals (data not shown). Measurements resulted in an accuracy better than 50 nm in all three dimensions: $x = 7 \pm 9$ nm; $y = 40 \pm 11$ nm; $z = 22 \pm 12$ nm.

**Acknowledgements**

We thank the Sanger Institute and Eric Schoenmakers (University Nijmegen) for providing BACs. We thank the Academic Medical Centre (University of Amsterdam), especially Mireille H.G. Indemans, for the confirmative sequencing of the BACs and the isolation of BAC DNA. We gratefully acknowledge Dr Erik M.M. Manders and Mr Wijnand Takkenberg from the Centre of Advanced Microscopy (University of Amsterdam) for technical support. Also, we acknowledge the helpful comments from Jens Odenheimer concerning data analysis. This work was supported by the European Commission as part of the 3DGENOME program: contract LSHG-CT-2003-503441.

**Supplementary Material**

Supplementary material can be found at: ..

**References**

1. Richmond TJ & Davey CA (2003) *Nature* 423, 145-150.
2. Schalch T, Duda S, Sargent DF, & Richmond TJ (2005) *Nature* 436, 138-141.
3. Woodcock CL & Dimitrov S (2001) *Curr Opin Genet Develop* 11, 130-135.
4. HorowitzScherer RA & Woodcock CL (2006) *Chromosoma* 115, 1-14.
5. Pederson T (1998) *J Mol Biol* 277, 147-159.
6. Cook PR (1999) *Science* 284, 1790-1795.
7. Bode J, Goetze S, Heng H, Krawetz SA, & Benham C (2003) *Chromosome Res* 11, 435-445.
8. Heng HH, Goetze S, Ye CJ, Liu G, Stevens JB, Bremer SW, Wykes SM, Bode J, & Krawetz SA (2004) *J Cell Sci* 117, 999-1008.
9. Cai ST, Lee CC, & KohwiShigematsu T (2006) *Nat Genet* 38, 1278-1288.
10. Kumar S, Allen GC, & Thompson WF (2006) *Trends Plant Sci* 11, 159-161.
11. Simonis M, Klous P, Splinter E, Moshkin Y, Willemsen R, de Wit E, van Steensel B, & de Laat W (2006) *Nat Genet* 38, 1348-1354.
12. Kurukuti S, Tiwari VK, Tavoosidana G, Pugacheva E, Murrell A, Zhao ZH, Lobanenkov V, Reik W, & Ohlsson R (2006) *Proc Nat Acad Sci USA* 103, 10684-10689.
13. Li QL, Barkess G, & Qian H (2006) *Trends Genet* 22, 197-202.
14. Fraser P (2006) *Curr Opin Genet Develop* 16, 490-495.
15. Pombo A, Jones E, Iborra FJ, Kimura H, Sugaya K, Cook PR, & Jackson DA (2000) *Crit Rev Eukaryot Gene Expr* 10, 21-29.
16. Tolhuis B, Palstra RJ, Splinter E, Grosveld F, & deLaat W (2002) *Mol Cell* 10, 1453-1465.
17. Volpi EV, Chevret E, Jones T, Vatcheva R, Williamson J, Beck S, Campbell RD, Goldsworthy M, Powis SH, Ragoussis J, *et al.* (2000) *J Cell Sci* 113, 1565-1576.
18. Sproul D, Gilbert N, & Bickmore WA (2005) *Nat Rev Genet* 6, 775-781.
19. Chambeyron S & Bickmore WA (2004) *Gene Develop* 18, 1119-1130.
20. Chambeyron S, DaSilva NR, Lawson KA, & Bickmore WA (2005) *Development* 132, 2215-2223.
21. Morey C, Da Silva NR, Perry P, & Bickmore WA (2007) *Development* 134, 909-919.
22. Van Driel R, Fransz PF, & Verschure PJ (2003) *J Cell Sci* 116, 4067-4075.
23. Lanctot C, Cheutin T, Cremer M, Cavalli G, & Cremer T (2007) *Nat Rev Genet* 8, 104-115.
24. Hahnfeldt P, Hearst JE, Brenner DJ, Sachs RK, & Hlatky LR (1993) *Proc Natl Acad Sci U S A* 90, 7854-7858.
25. Sachs RK, van den Engh G, Trask B, Yokota H, & Hearst JE (1995) *Proc Natl Acad Sci U S A* 92, 2710-2714.
26. Ostashevsky JY & Lange CS (1994) *J Biomol Struct Dyn* 11, 813-820.
27. Yokota H, Vandenengh G, Hearst JE, Sachs RK, & Trask BJ (1995) *J Cell Biol* 130, 1239-1249.
28. Vandenengh G, Sachs R, & Trask BJ (1992) *Science* 257, 1410-1412.
29. Munkel C, Eils R, Dietzel S, Zink D, Mehring C, Wedemann G, Cremer T, & Langowski J (1999) *J Mol Biol* 285, 1053-1065.
30. Caron H, van Schaik B, van der Mee M, Baas F, Riggins G, van Sluis P, Hermus MC, van Asperen R, Boon K, Voute PA, *et al.* (2001) *Science* 291, 1289-1292.

31. Versteeg R, van Schaik BD, van Batenburg MF, Roos M, Monajemi R, Caron H, Bussemaker HJ, & van Kampen AH (2003) *Genome Res* 13, 1998-2004.
32. Goetze S, Mateos-Langerak J, Gierman HJ, de Leeuw W, Giromus O, Indemans MHG, Koster J, Ondrej V, Versteeg R, & van Driel R (2007) *Mol Cell Biol*, MCB.00208-00207.
33. Cremer T & Cremer C (2001) *Nat Rev Genet* 2, 292-301.
34. De Gennis P (1979) *Scaling concepts in polymer physics* (Cornell University Press, Ithaca, N.Y.).
35. Liu B & Sachs RK (1997) *Bull Math Biol* 59, 325-337.
36. Bystricky K, Heun P, Gehlen L, Langowski J, & Gasser SM (2004) *Proc Natl Acad Sci U S A* 101, 16495-16500.
37. Yokota H, Singer MJ, vandenEngh GJ, & Trask BJ (1997) *Chromosome Res* 5, 157-166.
38. Gilbert N, Boyle S, Fiegler H, Woodfine K, Carter NP, & Bickmore WA (2004) *Cell* 118, 555-566.
39. Heun P, Laroche T, Shimada K, Furrer P, & Gasser SM (2001) *Science* 294, 2181-2186.
40. Gasser SM (2002) *Science* 296, 1412-1416.
41. Marshall WF, Straight A, Marko JF, Swedlow J, Dernburg A, Belmont A, Murray AW, Agard DA, & Sedat JW (1997) *Curr Biol* 7, 930-939.
42. Marshall WF (2002) *Curr Biol* 12, 185-192.
43. Chubb JR & Bickmore WA (2003) *Cell* 112, 403-406.
44. deLeeuw W, Verschure PJ, & vanLiere R (2006) *IEEE Transactions on Visualization and Computer Graphics* 12, 1251-1258.

**Figure legends**

Figure 1

**Fig. 1.** Distance measurements on the basis of the Human Transcriptome Map (HTM). **A.** Plots of the HTM showing the ridges and anti-ridges investigated on chromosomes 1 and 11. Every vertical line represents one gene and shows its median transcription over a window of 49 genes. Ridges are indicated by green boxes, anti-ridges by red ones. On the x-axis of the HTM, the giemsa metaphase banding pattern is indicated. (G bands in black, R bands in white). The arrows in the figure designate the regions where spatial distances between BAC probes were measured. The tail of each arrow indicates the position of the reference BAC, which was kept constant for a series of measurements. Loci at increasing genomic distances were selected in the direction of the arrowhead.. **B.** Plots showing average 3D physical distance (<R>) vs. genomic distance (g). Data points in green and red respectively correspond to the ridges and anti-ridges marked on chromosomes 1 and 11 in A. Error bars represent standard deviation. **C.** Plots showing average 3D physical distance vs. genomic distance for large genomic distances irrespective of transcriptional domain activity. Measurments were taken corresponding to the black arrows shown in figure 1A. Error bars represent standard deviation.

Figure 2

**Fig. 2**. Polymer models of chromatin folding.
The graphics show the experimimental data of figure 1 and those published by Yokota et al. (27) calculated with the random walk (RW), the self-avoiding walk (SAW) and the globular state (GS) polymer model. **A-C.** Plots showing $<R^2> / N_m^{2\nu}$ values against the genomic distance for the RW model ($\nu = 0.5$) (**A**), the SAW model ($\nu = 0.588$) (**B**) and the GS model ($\nu = 1/3$) (**C**). Green and red squares indicate the experimental values for the ridge and the anti-ridge of chromosome 1. Error bars represent the standard error. Blue pentagons represent 2D distance measurement values carried out by Yokota et al. (27) on a telomeric region (4p16.3) of chromosome 4. The linear regression lines in figures A-C illustrate the trend of the measured data sets assuming different polymer models. This regression line is horizontal only for the globular state model. **D.** Plots showing $<R^2> / N_m^{2\nu}$ vs. genomic distance values interpreted via the GS model ($\nu = 1/3$) for all aquired data sets. Green squares and triangles show the experimental values for the ridges on chromosomes 1 and 11, red squares and triangles code for the data measured for the anti-ridges of chromosomes 1 and 11. Black empty and filled circles represent long distance measurements on chromosomes 1 and 11. Error bars represent standard error. Blue rhombuses represent 2D measurements by Yokota et al. (27) along chromosome 4.

**Figure 1**

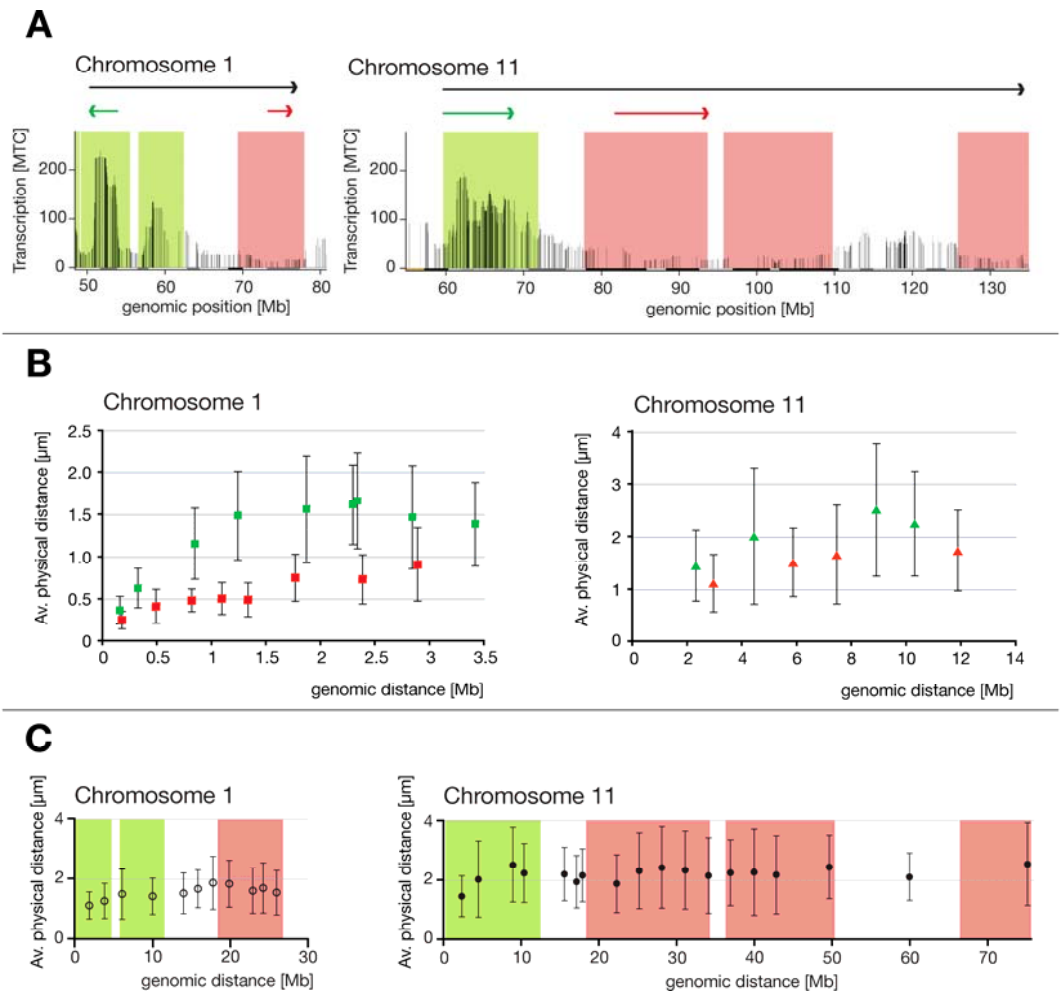

**Figure 2**

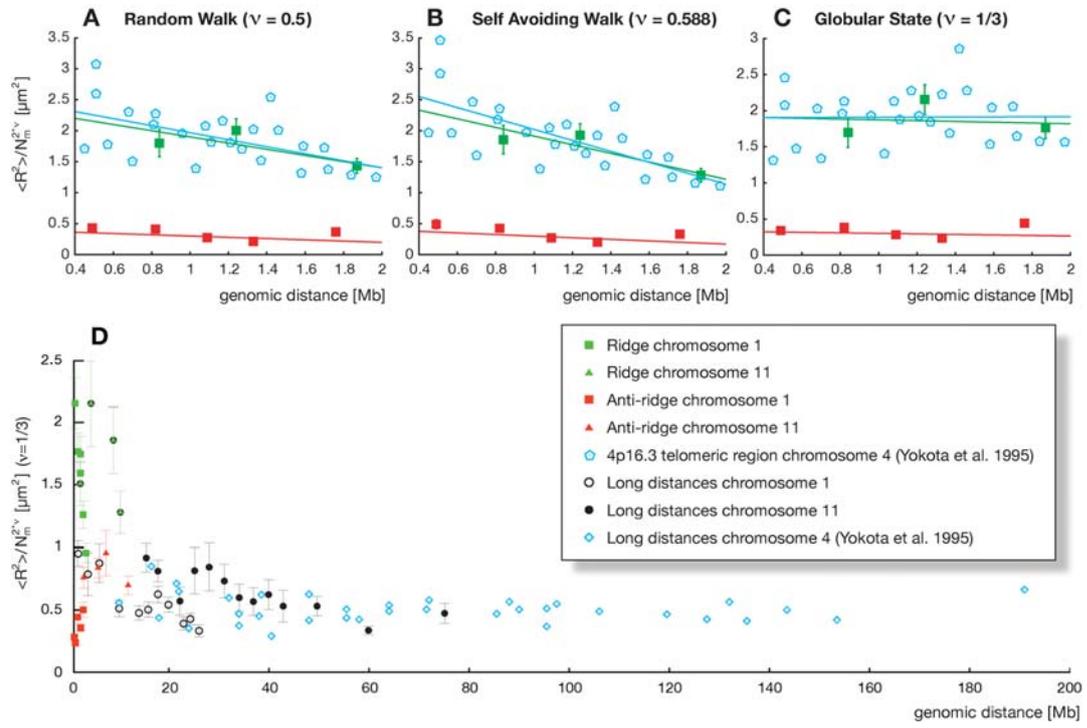